\documentclass[useAMS,usenatbib]{mn2e}
\usepackage{natbib,epsfig}

\title[Formation of the Double Pulsar]{The Formation of the Double Pulsar PSR~J0737$-$3039A/B}

\author[I.\ H.\ Stairs et al.]{I.\ H.\ Stairs$^{1}$\thanks{E-mail: stairs@astro.ubc.ca (IHS); thorsett@ucolick.org (SET); dewey@astro.ucsc.edu (RJD); mkramer@jb.man.ac.uk (MK)},
S.\ E.\ Thorsett$^{2}$, 
R.\ J.\ Dewey$^{2}$, 
M.\ Kramer$^{3}$,
C.\ A.\ McPhee$^{1}$\\
$^{1}$Department of Physics and Astronomy, University of British Columbia,
                Vancouver, BC V6T 1Z1, Canada\\
$^{2}$Department of Astronomy and Astrophysics, University of
                California, Santa Cruz, CA 95064, U.S.A.\\
$^{3}$University of Manchester, Jodrell Bank Observatory, Macclesfield, Cheshire SK11~9DL, U.K.}

\begin{document}

\date{Accepted 2006 August 29.  Received 2006 August 29; in original form 2006 July 19}

\pagerange{\pageref{firstpage}--\pageref{lastpage}} \pubyear{2006}

\maketitle

\begin{abstract}
Recent timing observations of the double pulsar J0737$-$3039A/B have
shown that its transverse velocity is extremely low, only 10\,km/s,
and nearly in the Plane of the Galaxy.  With this new information, we
rigorously re-examine the history and formation of this system,
determining estimates of the pre-supernova companion mass, supernova
kick and misalignment angle between the pre- and post-supernova
orbital planes.  We find that the progenitor to the recently formed
`B' pulsar was probably less than 2\,$M_{\odot}$, lending credence to
suggestions that this object may not have formed in a normal supernova
involving the collapse of an iron core.  At the same time, the
supernova kick was likely non-zero.  A comparison to the history of
the double-neutron-star binary B1534+12 suggests a range of possible
parameters for the progenitors of these systems, which should be taken
into account in future binary population syntheses and in predictions
of the rate and spatial distribution of short gamma-ray burst events.
\end{abstract}


\section{Introduction}

The double pulsar PSR~J0737$-$3039A/B \citep{bdp+03,lbk+04} is an
outstanding laboratory for tests of general relativity and of binary
evolution and supernova theories.  With its short orbital period,
$P_b=2.4$~hours, and moderate eccentricity, $e=0.088$, the
general-relativistic modifications to the Keplerian orbit are the
largest known.  Pulsars `A' and `B' have spin periods of
22.7\,msec and 2.7\,s, respectively; since both are observed as radio
pulsars, the mass ratio is obtained directly from pulsar timing,
leading to new theory-independent constraints on strong-field gravity
\citep{lbk+04, ksm+06}.

Such double-neutron-star (DNS) systems
are descended from high-mass X-ray binaries (HMXBs) in which both
stars are massive enough to undergo supernova (SN) explosions
\citep[e.g.,][]{tv03,dp03b}.  In brief, the J0737$-$3039A/B binary 
is thought to have begun as two main-sequence stars with masses of at
least 8\,$M_{\odot}$.  After a first mass transfer stage, the primary
formed a neutron star in a core-collapse SN.  The secondary evolved,
and matter was accreted by the neutron star in an HMXB phase.
Eventually, the secondary's envelope enlarged to meet the neutron
star, which spiraled in, ejecting the secondary's envelope.  Angular
momentum transferred to the neutron star spun it up to a period of a
few tens of milliseconds; it is now observed as the A pulsar.  Its
envelope expelled, the helium core of the secondary remained in a
circular orbit around the neutron star until a second supernova left
the B pulsar. The radio lifetime of the recycled A pulsar is far
longer than that of the high-magnetic-field B pulsar; this is the
reason most double-neutron star systems are observed with only the
recycled pulsar still active. We are fortunate to observe the double
pulsar during its relative youth.

The observed binary orbital elements and the space motion of the
binary preserve important details of the evolution, including the mass
lost in the second supernova (and hence the mass of the exploding
helium star), the size of the pre-supernova orbit, and any asymmetry
in the explosion itself. The lower limit on the mass of an exploding
helium core is of particular interest; evolutionary models suggest it
should be around 2.1\,$M_{\odot}$ \citep{nom84,hab86}.  Also of
interest is the misalignment angle $\delta$ between the first-born
NS's spin axis and the post-SN orbital angular momentum, equivalent to
the tilt between the pre- and post-SN orbital planes
\citep[e.g.,][]{wkk00}.  Constraints on all of the evolutionary
parameters are important for estimating the detectability of DNS
systems in pulsar surveys, and hence for estimates of the total DNS
birthrate and population. These quantities are of wide interest for
stellar evolution and nucleosynthesis, but particularly because
of their importance to event rate estimates for gravitational-wave
detectors \citep[e.g.,][]{phi91,kkl+04} and as possible progenitors
of short gamma-ray bursts \citep[GRBs;][]{pac86}.

Early attempts to constrain the pre-SN parameters and the kick
velocity of J0737$-$3039A/B have led to ambiguous results.  Shortly after
the discovery, analyses based on the orbital elements alone
\citep{dv04,wk04} suggested that moderate kick velocities (of 60\,km/s
or more) were needed, and that the pre-SN helium star was likely
low-mass and overflowing its Roche Lobe at the time of the explosion.
Later analyses by Willems et al.\ (2004, 2006) \nocite{wkh04,wkf+06}
used observational constraints on the transverse velocity of the
system, first $\sim 140\,$km/s from scintillation measurements
\citep{rkr+04} and later $<30$\,km/s from pulsar timing
\citep{kll+05}, to trace the motion of the binary back through the
gravitational potential of the Galaxy \citep{kg89} and derive
probability density functions for the most likely kick velocity,
pre-SN mass, and tilt angle.  In the most recent paper
\citep{wkf+06}, population synthesis models were also used to estimate
the most likely radial velocity of the binary (which is not directly
measurable); this helped to further restrict the allowed parameters.
Overall, these authors favor the standard formation scenario,
reasonably high kick velocities (70--180\,km/s) and pre-SN mass
$m_{2,i}$ of at least 2\,$M_{\odot}$, but acknowledge that for very
small transverse velocities ($\sim 10\,$km/s) lower-mass progenitors
are allowed or even favored.  Piran \& Shaviv (2005a,b)
\nocite{ps05,ps05b} have argued, based on the assumption that the
system will oscillate vertically about the Plane of the Galaxy, that
its current location close to the Plane implies that it must have a
low transverse velocity and have experienced a very small or no kick
at birth, indicating a pre-formation B mass less than 2\,$M_{\odot}$
and possibly an entirely different formation mechanism, such as the
electron-capture supernova proposed by \citet{pdl+05}.

\section{New results}

We are now in a position to reconcile these conflicting results.  Our
ongoing timing observations at the Parkes, Jodrell Bank Lovell and
Green Bank Telescopes have yielded a well-measured proper motion,
$\mu=4.2\pm0.4$\,mas/year, directed toward a celestial position angle
(North through East) of $308.4 \pm 6.5^{\circ}$, or a Galactic
position angle of $247.9 \pm 6.5^{\circ}$ \citep{ksm+06} -- that is,
its transverse motion is nearly parallel to the Plane of the Galaxy.
Our tentative detection of the timing parallax is consistent with the
adopted distance estimate of 520\,pc derived from the dispersion
measure \citep{cl02}.  Our measurements thus yield an extremely low
transverse velocity, only 10\,km/s, allowing us to derive a better
estimate of the history of the system's motion in the Galaxy.

To determine not just the limits of the allowed progenitor mass and
kick velocity, but also their most likely values, we adapt the
analysis that we previously used for the DNS PSR~B1534+12
\citep[hereafter TDS05]{tds05} to calculate posterior probability
density functions (pdfs) for several choices of priors on observable
or potentially observable parameters.

The unknown physical parameters are: the pre-supernova companion mass
$m_{2,i}$, the pre-supernova orbital separation $a_i$, the
3-dimensional supernova kick vector ${\bf V_k}$, the 3-dimensional
centre-of-mass velocity of the post-supernova system ${\bf V_{\rm
cm}}$, the `tilt' angle $\delta$ between the pre- and post-supernova
orbital planes, the angle $\Omega$ of the current binary's line of
nodes on the sky, and the sign of the cosine of the current system's
orbital inclination angle $i$ (note that $0^{\circ} < i <
180^{\circ}$ and that $\sin i$ is well measured through timing).  Two
components of the current 3-d velocity are now measured through
timing.  Two relations between $m_{2,i}$, $a_i$, and ${\bf V_k}$ can
be obtained through conservation of energy and momentum
\citep[e.g.][]{kal96,wkk00,wkf+06}.  \citet{wkf+06} derive the set of 
possible solutions for ${\bf V_k}$, weighting by a likelihood that
assumes a uniform prior on the magnitude $V_k$ and an isotropic
angular distribution.  They then convolve this function with their
pdfs for the system's radial velocity $V_r$, $\Omega$\footnote{The
Willems et al.\ (2006) definition of $\Omega$ differs from ours, but
as both studies assume uniform priors this is not very
important.} and probability that a given system will move to the
pulsar's current location.

We formulate the problem slightly differently,
using our insight from TDS05 that if $V_r$ (and hence ${\bf V_{\rm
cm}}$), $\delta$, $\Omega$ and $\cos i$ are measured or specified,
then at every potential birthsite in the Galaxy, the equations
connecting the remaining variables reduce to a simple quadratic
in $m_{2,i}$, with at most 2 possible real solutions.  This allows us
to assume prior distributions for $V_r$ and the potentially observable
$\delta$ and $\Omega$.  Our 3-d pdf may be written:
\begin{eqnarray*}
&& p(V_r, \Omega, \delta|D,I)\\
 &= &p(V_r, \Omega, \delta, {\bf V_{\rm pec}}, \cos i | m_1,m_{2,f},a_f,e_f,V_t,I) \\
& \propto & p(m_1,m_{2,f},a_f,e_f,V_t| V_r, \Omega, \delta, {\bf V_{\rm pec}}, \cos i ,I) \\
{} &{}& \times p(V_r, \Omega, \delta,{\bf V_{\rm pec}}, \cos i |I)
\end{eqnarray*}
\noindent where $m_1$ is the mass of pulsar A, $m_{2,f}$ is the 
mass of pulsar B, $a_f$ is the post-SN semi-major axis, and $e_f$ the
post-SN orbital eccentricity.  ${\bf V_{\rm pec}}$ represents the
3-dimensional peculiar velocity of the system before the second
explosion, and is treated as a nuisance parameter along with $\cos i$.
The likelihood $p(m_1,m_{2,f},a_f,e_f,V_t| V_r, \Omega, \delta, {\bf
V_{\rm pec}}, \cos i ,I)$ is simply 1 for each acceptable real
positive-mass solution of the quadratic equation for that choice of
parameters, and 0 otherwise.

Our analysis proceeds as follows:  For each of a large number of
trials, we pick a radial velocity $V_r$ from a prior distribution.  As
discussed by \citet{ps06} and recognized by \citet{wkf+06}, the choice
of this prior has important effects on the final pdfs.  For the
purposes of illustration, we investigate two prior distributions: 1)
gaussian in $V_r$ with a dispersion 200\,km/s \citep{wkf+06}, and 2)
$V_r = V_t / \tan \theta$, where $\cos \theta$ is chosen uniformly
between $-1$ and $+1$.  The second prior is the one preferred by Piran
\& Shaviv \nocite{ps05b,ps06} and is arguably the more logical
prior to use\footnote{We note that the expected radial velocity of the
pulsar's LSR to that of the Sun is very small, on the order of
5\,km/s.  Given the uncertainties in, for example, the gravitational
potential of the Galaxy, we consider it appropriate to discuss radial
velocities centred on 0\,km/s.}.  We consider the first prior to be
extremely conservative, in that its dispersion is much larger than the
observed transverse velocities of DNS systems.  We sample the proper
motion from gaussian distributions using the measured uncertainties on
magnitude and direction, and the pulsar distance assuming a 20\%
gaussian uncertainty.

We then follow the motion of the binary system back in time through
the Galaxy, similarly to the procedure in \citet{wkf+06} and TDS05,
incorporating the Sun's position and peculiar motion.  We accept any
position with $|z|<50\,$pc and galactocentric radius $R < 15\,$kpc as
a possible birth site for the B pulsar.  We integrate back in time the
orbital eccentricity and semi-major axis, according to the equations
of \citet{pet64}, testing ages up to 100\,Myr.

At each birth site, 1000 trial sets of parameters are selected. The
peculiar velocity $\bf V_{\rm pec}$ is assumed to be a Maxwellian with
a 1-dimensional dispersion of 12\,km/s, based on the proper motions of
Be/X-ray binaries \citep{ci98,vpbk00}, while the $\cos i < 0$ and
$\cos i > 0$ cases are treated as equally likely.  We draw each of
$\Omega$ and $\delta$ from uniform distributions: $0^{\circ} <
\Omega < 360^{\circ}$ and $0^{\circ} < \delta < 180^{\circ}$.
The angle $\delta$ is potentially observable but currently unknown,
although the lack of profile shape changes in A \citep{mkp+05} implies
that it is probably small.  The angle $\Omega$ can in principle be
estimated from scintillation \citep{cmr+05} but its value depends
strongly on the modeling of the anisotropy of the interstellar medium
and is not currently well constrained (Coles \& Rickett, private
communication). For
each set of parameters, we construct the quadratic equation and
determine whether there are solutions.  We record progenitor and
kick parameters and construct pdfs via histograms. 

An important point neglected by both Willems et al.\ and Piran \&
Shaviv is that a neutron star's mass contains a large
negative contribution from gravitational self-energy, meaning that the
minimum progenitor mass must be equal to the current neutron-star mass
plus the magnitude of the binding energy $E_{\rm B}$.  This quantity
can be estimated, to within about 20\%, as $E_{\rm B} \simeq 0.084(M/M_{\odot})^2 M_{\odot}$
\citep{ly89,lp01}.  The majority of plausible equations of
state predict a very slightly lower binding energy for a
1.25\,$M_{\odot}$ neutron star.  The net implication is that a minimum
(baryonic) mass of about 1.37\,$M_{\odot}$ is required for the
progenitor of B, and we consider only solutions which yield
progenitor masses above this limit.

To place the double pulsar in context, we compare it to the only other
DNS with strong constraints on the progenitor, PSR~B1534+12 (TDS05).
(Recent proper motion measurements for PSR~B1913+16 will necessitate a
revised set of kick constraints for that pulsar \citep{nwt06}.)  For a
fair comparison, we have revisited our analysis of B1534+12,
calculating pdfs as for J0737$-$3039A/B, while allowing for progenitor
masses down to 1.48\,$M_{\odot}$ and {\it not} using the scintillation
constraint on $\Omega$ \citep{bplw02} in case anisotropy systematics
also affect this measurement.  Because of the strong misalignment
angle constraint, fewer sets of trial parameters are tested per birth
site, while more separate radial velocity trials are used.

\section{Discussion}

\begin{figure}
\epsfig{file=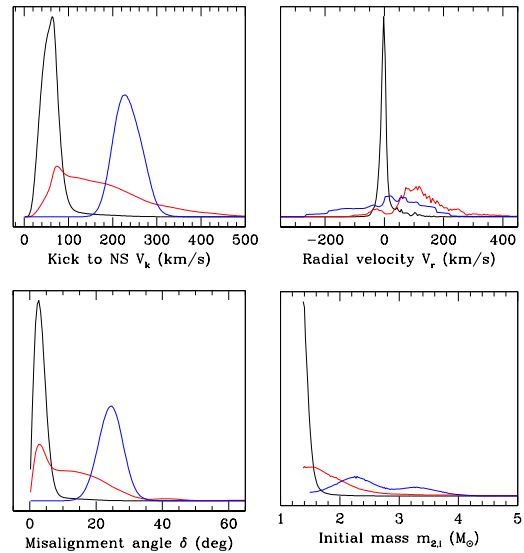,width=8cm}
\caption{Normalized posterior pdfs for DNS progenitors.  In each panel,
the red and black lines represent PSR~J0737$-$3039A/B progenitors,
with the red line corresponding to the gaussian $V_r$ prior with
dispersion $200\,$km/s, and black to the $V_r$ prior which is uniform
in the cosine of the angle between the velocity vector and our line of
sight.  The blue lines show pdfs for progenitors of PSR~B1534+12, for
the gaussian $V_r$ prior with dispersion $200\,$km/s.  Only parameter
ranges with significantly non-zero pdf values are
shown. Note the association of large $\delta$ with large kicks, 
as expected.\label{fig:results}}
\end{figure}

Because of the constraint equations used, each point in the (${\bf
V_k}$, $m_{2,i}$, $a_i$) parameter space corresponds to a point in the
($V_r$, $\delta$, $\Omega$) parameter space.  Thus we have effectively
constructed the posterior pdfs for the (${\bf V_k}$, $m_{2,i}$, $a_i$)
parameters as well, and may marginalize to derive confidence ranges on
all the parameters.  For both J0737$-$3039A/B and B1534+12, our pdfs
for the physically interesting quantities are plotted in
Figure~\ref{fig:results} and confidence ranges are summarized in
Table~\ref{tab:conf}.  We recognize, as pointed out by \citet{wkf+06},
that projection effects are neglected in these 1-d pdfs.  The most
important such effect for J0737$-$3039 is the association of
large-${\bf V_k}$, large-$m_{2,i}$ and large-$\delta$ solutions
exclusively with the large $V_r$ ($\stackrel{>}{\sim} 100\,$km/s)
progenitors accessible only with the gaussian $V_r$ prior.  Because
these imply that the current velocity makes a very small angle to our
line of sight, we regard them as inherently less plausible.

\begin{table*}
 \centering
 \begin{minipage}{140mm}
\caption{Confidence ranges on progenitor parameters\label{tab:conf}}
\begin{tabular}{@{}lccccccccc@{}}
\hline
\multicolumn{1}{c}{$V_r$ prior}
& \multicolumn{3}{c}{Median likelihood value}
& \multicolumn{3}{c}{68\% confidence interval}
& \multicolumn{3}{c}{95\% confidence interval} \\
{}
& {$V_k$}
& {$m_{2,i}$}
& {$\delta$} 
& {$V_k$}
& {$m_{2,i}$}
& {$\delta$} 
& {$V_k$}
& {$m_{2,i}$}
& {$\delta$} \\
{}
& {(km/s)}
& {($M_{\odot}$)}
& {($^{\circ}$)} 
& {(km/s)}
& {($M_{\odot}$)}
& {($^{\circ}$)} 
& {(km/s)}
& {($M_{\odot}$)}
& {($^{\circ}$)} \\
\hline
\multicolumn{10}{l}{J0737$-$3039A/B} \\
Gaussian & 165 & 1.80 & 12.0 & 80--305 & 1.50--2.40 & 3.0--24.5 & 45--1005 & 1.37--4.00 & 1.0--102.5 \\
Uniform-in-$\cos$ & 60 & 1.45 & 3.5 & 40--80 & 1.37--1.55 & 1.5--5.5 & 20--140 & 1.37--1.80 & 0.5--11.0 \\
 & & & & & & & & &\\
\multicolumn{10}{l}{B1534+12} \\
Gaussian & 235 & 2.45 & 24.5 & 200--270 & 2.00--3.35 & 20.5--28.5 & 175--305 & 1.60--3.90 & 16.5--32.5 \\
\hline
\end{tabular}
\end{minipage}
\end{table*}

We can draw several conclusions about the J0737$-$3039B progenitor and
the supernova kick.  First, the explosion was asymmetric, but not
extremely so.  The $V_k$ pdf extends down toward $0\,$km/s for both
$V_r$ priors, but a non-zero kick is strongly favoured.  That the
supernova explosion was most likely asymmetric is bolstered by the B
pulsar's spin-orbit misalignment observed through long-term changes in
the B profile \citep{bpm+05} and derived through modeling of the A
eclipses \citep{lt05,lyu05}.  For the gaussian $V_r$ prior, the kick
tends to be directed out of the plane of the pre-SN orbit and away
from the pre-SN progenitor orbital velocity.  For the
uniform-direction $V_r$ prior, the kick is directed nearly randomly in
the plane perpendicular to the pre-SN progenitor orbital velocity.

The only likely radial velocities are either slightly negative or else
positive.  This reflects the fact that large negative velocities
(relative to our current reference frame) imply that the system must
have been born in the outer reaches of the Galaxy, but there are far
fewer potential birth sites at large Galactic radii.

Low, even very low ($ < 10^{\circ}$), misalignment angles are
predicted between A's spin axis and the orbital angular momentum, in
excellent agreement with the observed lack of profile variations
\citep{mkp+05,ksm+06}.  

The immediate B progenitor was probably less than about
2\,$M_{\odot}$.  Since the pre-SN orbital period was comparable to the
current 0.1-day value, the low mass is likely due to significant mass
loss accompanying the orbital shrinkage. \citet{dpsv02} and
\citet{dp03b} trace the histories of He-star/NS binaries,
showing that suitable parameters can be obtained from a range of
starting points.  A difficulty is that the lowest-mass systems are
expected to undergo common-envelope evolution and spiral-in, resulting
in extremely tightly-bound or even merged systems (\citet{dp03b}, but
see \citet{ibk+03}).  Dewi \& Pols note, however, that the relative
time-scales of the spiral-in and supernova explosion are not well
known, and an explosion might occur before the spiral-in is complete.

The low-mass B progenitor is consistent with either an
electron-capture collapse of the ONeMg core of a low-mass He-star
\citep{pdl+05} or the collapse of a low-mass iron core.  Either
might occur sufficiently fast to prevent the development of large
asymmetries, resulting in the small kick \citep{prps02,plp+04}.
Estimates of the time-scales for the onset of the two types of
explosions are similar \citep{dp03b} and appear too long relative to
the spiral-in time.  Better modeling of the two types of explosions
may ultimately allow us to distinguish between the two on the basis of
the required short time-scale.

PSR~B1534+12 presents an interesting contrast.  Our derived pdfs are
fully compatible with the parameter space allowed in TDS05; in
particular the kick velocity is likely in the range 200--270\,km/s
while the progenitor companion mass was in the range
2.0--3.35\,$M_{\odot}$.  We report the confidence intervals only for
the gaussian (dispersion 200\,km/s) $V_r$ prior in
Table~\ref{tab:conf}; the uniform-in-cos prior gives similar results.
We note that the most important constraint in this case is the
measured misalignment angle \citep[$25\pm 4^{\circ}$;][]{sta04}, whose
only model-dependence is on the reasonable assumption that the
rotating vector model \citep{rc69a} describes the linear polarization
position-angle swing of this pulsar \citep[c.f.][]{dpp05}.  Overall,
the parameters of the B1534+12 progenitor fit quite well with a more
`standard' evolutionary picture that involves He-star Roche-lobe
overflow but not a late spiral-in phase.  It is possible that the
longer orbital period and/or higher mass of the B1534+12 progenitor
permitted a larger core to develop before the SN explosion, leading to
a larger kick; we note that the second-formed NS in B1534+12 is
0.1\,$M_{\odot}$ more massive than J0737$-$3039B.  Also of interest is
that \citet{ikb06} find PSR~B1913+16, with the largest known
second-formed NS mass, cannot be formed in population syntheses with
small kick velocities.

The contrast between these systems urges caution in discussing
expectations based on low kicks.  In the last several years, small
($\sigma \sim 50\,$km/s) kicks have been proposed to be nearly
universal for nascent neutron stars processed in relatively close
binary systems, due to the early loss of the NS progenitor's envelope
and subsequent formation of a low-mass core which in turn provides a
small kick upon collapse \citep[e.g.,][]{plp+04}.  This hypothesis has
been used to explain the existence of long-period, low-eccentricity
high-mass X-ray binaries \citep{prps02}, the retention of large
numbers of neutron stars in globular clusters \citep{prps02,plp+04},
the apparent correlation between the orbital eccentricity and the
recycled pulsar spin period in DNS systems \citep{fkl+05,dpp05} and
the apparent dearth of isolated mildly recycled pulsars (`failed
DNSs') ejected from unbinding second SN explosions \citep{dpp05}, as
well as to predict larger numbers of DNS mergers than are inferred
from the observed set of objects \citep{prps02}.  Our derived
progenitor parameters for J0737$-$3039A/B certainly imply that some
fraction of DNS progenitors will be low-mass and experience small
kicks, and hence that each of these explanations is plausible at some
level.  However, the much larger kick needed for B1534+12, which is
also a short-period (10h) binary, indicates that a range of parameters
must be considered in all of these arguments.  Noting the additional
rough correlation between the orbital eccentricity and the mass of the
second-formed compact object in the six systems for which the masses
are well-determined (PSRs J1756$-$2251, J0737$-$3039, J1141$-$6545,
B2303+46, B1534+12 and B1913+16; see also \citet{fau04}), we speculate
that the magnitude of the kick may depend quite sensitively on the
size of the collapsing core.  This speculation is not supported by the
most recent 2-D numerical simulations of hydrodynamic instabilities
during SN explosions \citep{skjm06}, which suggest that core mass and
NS velocity are not correlated, but more work will be required on 3-D
simulations and on other kick models. At the same time, careful
population synthesis, incorporating initial mass functions and
preferably an improved understanding of the supernova vs.\
orbital-evolution time-scales, will be required to determine whether
this correlation or the (likely related) spin-period--eccentricity
relation can truly be reproduced by a mass-dependent kick.

As pointed out by \citet{pdl+05}, the low space velocity of
J0737$-$3039 may provide a further probe of its long-term evolutionary
history.  A competing model to the `standard' DNS evolutionary model
outlined in Section~1 is the `double-core' scenario
\citep{bro95,bb98b} in which the He cores of two main-sequence stars
of nearly equal mass undergo a spiral-in through the envelopes of both
stars simultaneously.  Thus at the time of the first SN, the secondary
star is a (low-mass) He star rather than a massive main-sequence star,
and the systemic velocity after the first supernova explosion would be
expected to be higher than in the standard model.  The observed low
space velocity therefore makes the double-core scenario less likely.
If SN kicks are in fact lower in He stars processed in binaries, this
argument is slightly weakened; however we note that under the
mass-dependent kick hypothesis, since A's mass is close to that of the
companion in B1534+12, its kick velocity might well have been large.
 
Finally, our low-velocity progenitor for J0737$-$3039A/B may have
implications for the nature of short GRBs.  In the popular DNS
inspiral models \citep{pac86}, a poor correlation between active star
formation and GRB activity can be a natural consequence of the time
delay between birth and inspiral and the space velocity of the
post-supernova binary.  The few detected counterparts to short GRBs
place these objects in the outskirts of a broad variety of
galaxies without active current star formation
\citep[e.g.,][]{pbc+06}.  Allowing for large DNS kicks, this is 
consistent with the production of the progenitors
during normal galactic star formation \citep{bpb+06}.  However, if smaller
DNS kicks are assumed, then it may also be necessary to produce a
significant fraction of the progenitors in globular clusters in the
galaxy haloes \citep{gpm06}.  Our low velocity for J0737$-$3039A/B
tends to support the latter scenario, but also suggests that similar
systems could be a source of GRBs near the Earth, with
possible implications for our terrestrial environment
\citep{tho95b,tmj+05}.

\section*{Acknowledgments}
IHS holds an NSERC UFA and pulsar research at UBC is supported by a
Discovery Grant.
RJD and SET are supported by
the NSF under grants AST 0506453 and AST 0098343.
CAM holds an NSERC USRA.  We thank Maxim Lyutikov for helpful conversations.
This work was carried out on a computer cluster funded by a 
CFI New Opportunities grant to IHS and M.\ Berciu.

\end{document}